\documentclass[reprint,amsmath,amssymb,aps,pra,]{revtex4-1}
\pdfoutput=1
\usepackage{graphicx}
\usepackage{dcolumn}
\usepackage{bm}

  \newcommand{\ket}[1]{\left| #1 \right>} 
  \newcommand{\bra}[1]{\left< #1 \right|} 
  \newcommand{\braket}[2]{\left< #1 \vphantom{#2} \right| \left. #2 \vphantom{#1} \right>} 
  
\begin{document}

\preprint{APS/123-QED}

\title{ Quantum walks on graphs representing the firing patterns of a quantum neural network}

\author{Maria Schuld}
\email{schuld@ukzn.ac.za}
\author{Ilya Sinayskiy}
\author{Francesco Petruccione}
\affiliation{Quantum Research Group, School of Chemistry and Physics, University of KwaZulu-Natal Durban, KwaZulu-Natal, 4001, South Africa\\
and National Institute for Theoretical Physics (NITheP), KwaZulu-Natal, 4001, South Africa }

\date{\today}
            
\begin{abstract}
Quantum walks have been shown to be fruitful tools in analysing the dynamic properties of quantum systems. This article proposes to use quantum walks as an approach to Quantum Neural Networks (QNNs). QNNs replace binary McCulloch-Pitts neurons with a qubit in order to use the advantages of quantum computing in neural networks. A quantum walk on the firing states of such a QNN is supposed to simulate central properties of the dynamics of classical neural networks, such as associative memory. It is shown that a biased discrete Hadamard walk derived from the updating process of a biological neuron does not lead to a unitary walk. However, a Stochastic Quantum Walk between the global firing states of a QNN can be constructed and it is shown that it contains the feature of associative memory. The quantum contribution to the walk accounts for a modest speed-up in some regimes.\bigbreak
\begin{description}
\item[PACS numbers] 03.67.Lx, 03.65.Yz, 87.18.Sn
\end{description}
\end{abstract}

\pacs{Valid PACS appear here}

\maketitle

\section{Introduction}

Quantum walks, the quantum equivalent of classical random walks, became a booming research field in the last decade \citep{kempe03, kendon07, venegas12}. Based on the theory of Markov chains, classical random walks study the evolution of the probability distribution of an abstract walker's position on a graph. The positions or vertices of the graph are connected by edges symbolising transition probabilities. In each step, the walker makes a random decision (often described by a coin toss) to which adjacent position to jump. Quantum walks, in which the walker's position on the graph can be a superposition and the decision process is simulated by a `quantum coin' such as the Hadamard transformation, show a surprisingly different behaviour to classical random walks. Quantum walks have been formulated as discrete \citep{aharonov01} or continuous \citep{childs02} walks, and led to new versions such as the semi-classical Stochastic Quantum Walk \citep{whitfield10} or the Open Quantum Walk \citep{attal12}. \\

The potential of quantum walks is based on their fruitful application for quantum computing just like classical walks lead to efficient classical algorithms \citep{moore02, ambainis03}. An important application of quantum walks has been found in the newly emerging field of quantum biology \citep{ball11}. Evidence suggests that photosynthetic plants use nontrivial quantum effects such as superposition and interference for energy transport in their  light-harvesting complexes \citep{engel07, panitchayangkoon10}. The trajectory of an excitation `jumping' between molecular `sites' from the antenna to the reaction center can thereby be modeled with the formalism of quantum walks \citep{ishizaki09, hoyer10, mohseni08, rebentrost09}. \\

The success story of quantum biology is a motivation to reaccess questions of quantum dynamics in another important biological system: the brain. In 2006, Christof Koch and Klaus Hepp wrote in their  \textit{Nature} contribution entitled `Quantum Mechanics in the Brain' \citep{koch06}: ``The critical question [...] is whether any components of the nervous system - a $300^{\circ}$ Kelvin tissue strongly coupled to its environment - display macroscopic quantum behaviours, such as quantum entanglement, that are key to the brain's function.'' Besides a number of controversial theories on the `quantum brain' \citep{hameroff13, freeman08}, there has been no evidence for nontrivial quantum effects in the nervous system yet. On the contrary, the macroscopic nature of signal transmission between neural cells seems to make quantum coherence impossible \citep{tegmark00}. However, the intersection of neuroscience and quantum physics has been accessed from the perspective of computational science. In the last two decades, various Quantum Neural Network (QNN) models \citep{andrecut02, altaisky01,gupta01, behrman02, fei03, zhou12, oliveira08, toth00} have been proposed. Although not claiming to be realistic quantum models of biological neural networks, these proposals explore alternative ways of computation using both the advantages of quantum computing and neural computing.\\

This article follows the perspective of QNN research and investigates a new approach to introduce quantum physics into neural networks by making use of the theory of quantum walks. The sites of the quantum walk symbolise the firing patterns of a neural network consisting of simplified binary neurons with the states `active' and `resting'. A firing pattern is given by a binary string encoding which neuron of a network is firing and which is resting. The current position of the walker represents the network's firing state. A \textit{quantum} walker is of course able to be in a superposition of firing patterns. We show that a discrete quantum walk, in which a Hadamard-like biased coin is successively flipped to decide on the firing state of single neurons clashes with the framework of unitary quantum walks. To simulate a neural network's dissipative dynamics, we therefore need to focus on quantum walks that incorporate decoherence. A continuous Stochastic Quantum Walk on the hypercube, obtaining basic features of the brain's property of associative memory (i.e. retrieving a memorised pattern upon an imperfect initial pattern), is implemented and analysed. It can be shown that under certain conditions, the quantum part of the walk accounts for a modest speed-up of the walk. These results serve as an example of the application of quantum walks to obtain specific dynamics of a quantum system.\\

The paper has the following structure: Section II and III very briefly introduce into the necessary theoretical background of quantum walks as well as Quantum Neural Networks. Section IV gives an idea of how quantum walks can be constructed in the context of QNNs, and explains the reason for the failure of the most intuitive way. A more mature approach is presented. The conclusion (Section V) offers a discussion including a way forward for the use of quantum walks in QNN research.

\section{A brief introduction to quantum walks}
\begin{figure}[t]
  \centering    \includegraphics[width=0.25\textwidth]{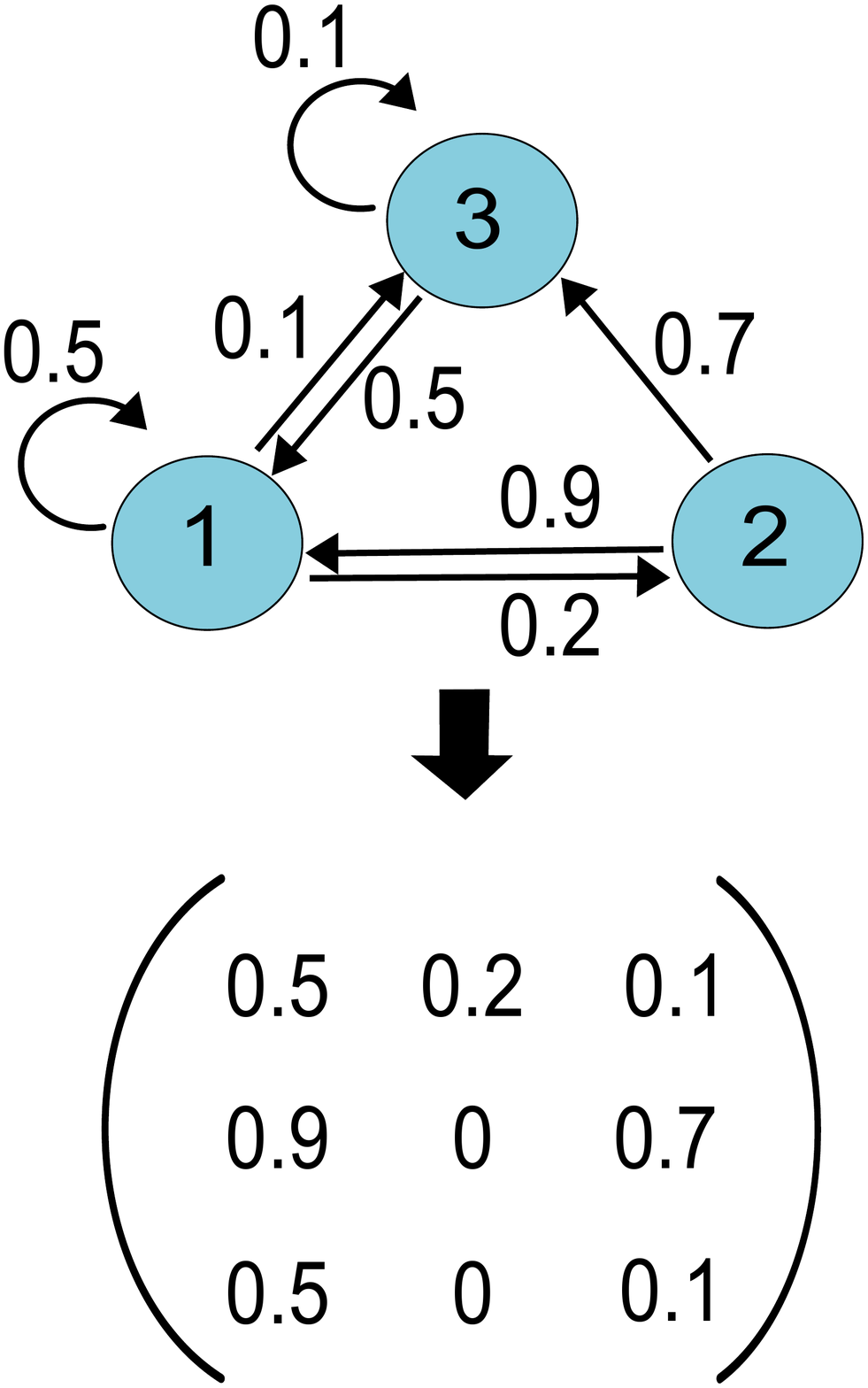} \caption{(Colour online) Any weighed graph defines a Markov chain represented by a stochastic matrix.} \label{markovchain}
\end{figure}

On any graph of $n$ vertices a Markov chain can be defined. A Markov chain is a sequence of events that is governed by a stochastic process in which the results of a timestep only depend on the results of the previous timestep \citep{feller08}. Markov chains are described by a stochastic matrix $ M(n\times n, \mathbb{R})$ with $\sum^n_{j = 1} m_{ij} = 1$ and entries $m_{ij}$ representing the weight of the directed edge going from vertex $i$ to $j$ (see Fig. \ref{markovchain}). These weights can be interpreted as a transition rate from site $i$ to $j$. Repeatedly applying $M$ to a $n$-dimensional stochastic vector $\vec{\pi}$ with $\sum^n_{l=1} \pi_{l}=1$ evolves an initial probability distribution through discrete time steps. The probability of being at vertex $i$ changes according to \citep{childs02}
\begin{equation} \frac{d\pi_i}{dt} = -\sum \limits_j  M_{ij}\pi_j(t). \label{machain} \end{equation}
Markov chains on regular undirected graphs result in a stationary probability distribution $\pi_s$ which is independent of the initial state \citep{aharonov01}. The time it takes to reach the stable distribution is called the \textit{mixing time}.\\

Markov chains with equal probability to jump from site $i$ to any of the $d$ sites adjacent to $i$ are also known as random walks on a graph \footnote{Some authors extend the random walk model to biased probabilities  \citep{mackay02, kendon07, stefanak09, ribeiro04}. These so called `biased random walks' are nothing else than Markov chains introduced here. We will use the more general definition.}. Random walks are based on the idea of an abstract walker who in each step tosses a $d$-dimensional coin to choose one of $d$ possible directions at random. Random walks have been proven to be powerful tools in constructing efficient algorithms in computer science (for references see \citep{moore02}). \\ 

In the quantum equivalent of random walks, a quantum walker walks between sites by changing its position state $\ket{x} \in \{\ket{1},..., \ket{n}\}$. The difference to classical walks is twofold: First, the various paths are realised in a superposition and thus interfere with one another, and a measurement `collapses' the paths into a current position. Second, the dynamics have to preserve the squared amplitude vector instead of the stochastic vector to preserve the total probability. This means that the evolution has to be unitary, or in the most general case of an open system, a completely positive trace preserving map \citep{aharonov02, nielsen10}. \\

The unitarity of quantum walks furthermore implies that the evolution is reversible. Quantum walks therefore do not have a stationary probability distribution $\pi_s$ as classical random walks do. However, it turns out that taking an average over the probability distribution over states $\ket{x}$,
\begin{equation} \bar{P}_T(x) = \frac{1}{T} \sum^{T-1}_{t=0} | \braket{x}{\psi(t)}|^2, \label{LD}\end{equation}
 leads to a stable distribution $\bar{P}_s$ \citep{aharonov01}. Quantum walks received a fair amount of attention and have been the topic of extensive reviews, books and attempts of implementations \citep{kendon07, kempe03, venegas12, wang13, travaglione02}. The reasons are that first, quantum walks show markedly different features than their classical counterparts. Second, quantum walks were in some cases able to outperform classical walks \citep{kempe02, childs02, aharonov02, ambainis03, shenvi03}.\\

Two versions of quantum walks were established and exist in parallel: the discrete \citep{aharonov01} and continuous time quantum walk \citep{farhi98, childs02}. The bridge between the two has finally been shown in \citep{strauch06}. An important development was also the exploration of decoherence in quantum walks \citep{kendon07}. Recently, an interesting version of the continuous quantum walk with decoherence has been introduced \citep{whitfield10}. So called Stochastic Quantum Walks obey a Gorini-Kossakowski-Lindblad-Sudarshan (GKLS) type master equation \citep{lindblad76,gorini78}  that consists of a coherent as well as an incoherent part. These three versions will be important in the application further down and shall therefore be briefly presented.\\

\textit{Discrete quantum walks}\\
In a discrete quantum walk, the `walker' is associated with a wave function describing a quantum system with states $\ket{\psi} = \ket{c} \otimes \ket{i} \in \mathcal{H}_c \otimes \mathcal{H}_i$. The Hilbert space $\mathcal{H}_i$ has a basis$\{\ket{0},..., \ket{n}\}$ ($n$ may be countable infinite) that represent sites or vertices on which the walk takes place. The Hilbert space $\mathcal{H}_c$ with basis $\ket{1},..., \ket{d_i}$ is a `coin' space that denotes the current state of a coin `tossed' to decide which direction to take next. Note that usually only regular graphs are considered and $d$ is independent of the current position $\ket{i}$. The discrete walk then follows two substeps in each step $t\rightarrow t+1$, executed by a coin and a shift operator: 
\[\ket{\psi_{t+1}} = \hat{S}(\hat{C} \otimes 1)\ket{\psi_t}.\]
First, the coin is `tossed' by applying $\hat{C}$ to the coin space. By that, the coin state is put into a superposition. Second, the conditional shift operator $\hat{S}$ shifts the state to the $r$'th adjacent site if the outcome of the coin is $\ket{r}$ ($r \in \{1,...,d\}$). \\

The most well-known coin is the Hadamard transformation that works on the two-dimensional basis $\{\ket{a}, \ket{b}\}$ as 
\begin{equation}\hat{H} \ket{a} = \frac{1}{\sqrt{2}} (\ket{a} + \ket{b}), \; \; \hat{H} \ket{b} = \frac{1}{\sqrt{2}} (\ket{b} - \ket{a}). \label{had}\end{equation}
The minus in the second equation indicates that even the Hadamard transformation is not completely unbiased and denotes at the same time the fundamental difference to classical walks, as it is the source of interference \footnote{The reason why a separate coin space is introduced is that the unitarity condition is not fulfilled by most examples of walks (a detailed argumentation can be found in \citep{ambainis03}).}.\\ 

\textit{Continous quantum walks}\\
In 2001, Andrew Childs, Edward Farhi and Sam Gutman published a continuous version of the quantum walk. Their idea is based on the equivalence between Eq. (\ref{machain}) and the Schr\"odinger equation for state $\ket{\psi (t)}$
\begin{equation} \imath \frac{d}{dt} \braket{i}{\psi(t)} = \sum \limits_j \bra{i}H\ket{j} \braket{j}{\psi(t)} \label{SE} \end{equation}
While  Eq. (\ref{machain}) preserves  $\sum^n_{l=1} \pi_{l}=1$,  the Schr\"odinger equation (\ref{SE}) makes sure that $\sum_i |\braket{i}{\psi(t)}|^2 = 1$ is fulfilled. The difference between the two evolutions is simply the imaginary unit in the latter \citep{childs02}. Comparing both equations, one can see that the stochastic matrix $M$ of the classical Markov chain is replaced by the Hamiltonian $H$ of the quantum system. $H$ consequently equals the weighed adjacency matrix of the graph. To obtain a hermitian and thus symmetric operator (as the entries are real numbers), the graph needs to be undirected.\\

\textit{Stochastic Quantum Walks}\\
A number of contributions investigate what happens if decoherence is introduced into quantum walks \citep{kendon07, brun03}. Decoherence destroys the quantum property of coherent states and drives the dynamics into the classical regime. In some cases this can lead to preferred dynamics \citep{kendon07}. An interesting proposal for a decohered continuous quantum walk has recently been introduced \citep{whitfield10}. A so called Stochastic Quantum Walk (SQW) evolves a density matrix according to the GKLS master equation \citep{lindblad76,gorini78}
\begin{equation}  \dot{\rho} =- \imath \kappa [H, \rho] - \gamma \sum_k \left( \frac{1}{2} L_k^{\dagger} L_k \rho + \frac{1}{2}\rho L_k^{\dagger} L_k  -  L_k\rho L_k^{\dagger} \right). \label{lindblad} \end{equation}
Note that here and in the following, $\hbar$ is set to one. The parameters $\kappa$ and $\gamma$ define the influence of the two terms on the right side. The sum term describes the stochastic evolution, in which $L_k$ denote Lindblad jump operators that decohere the quantum state. The first term is the usual Schr\"odinger quantum evolution as known from the von Neumann equation. The Hamiltonian represents a (weighed) adjacency matrix as in the continuous walk given in Eq. (\ref{SE}). In this fashion, the ``evolution among vertices happens through coherences developed by a Hamiltonian'' \citep{whitfield10} while the system is constantly exposed to decoherence. Thus, both advantages of dissipation and coherence can be used and transitions from one to the other studied. We will implement a version of the Stochastic Quantum Walk in Section IVb.

\section{Neural Networks and the concept of the `quron'}

To get an understanding of what is meant by the `firing patterns of a QNN', we need to briefly introduce into some fundamentals of computational neuroscience as well as the basic concept of a QNN.\\

Neural networks are computational systems inspired by the biological neural networks forming our brain. The brain is believed to compute information by carrying electric signals, so called \textit{action potentials}, along the membranes of interconnected neural cells \citep{purves08, dayan01}. The algorithmic dynamics of biological neural networks are defined by the connection strengths between neurons as well as by the activation function of a neuron due to the input signals from other neurons. Information is encoded in the global firing pattern of the neural network.\\ 

\begin{figure}[t]
  \centering    \includegraphics[width=0.45\textwidth]{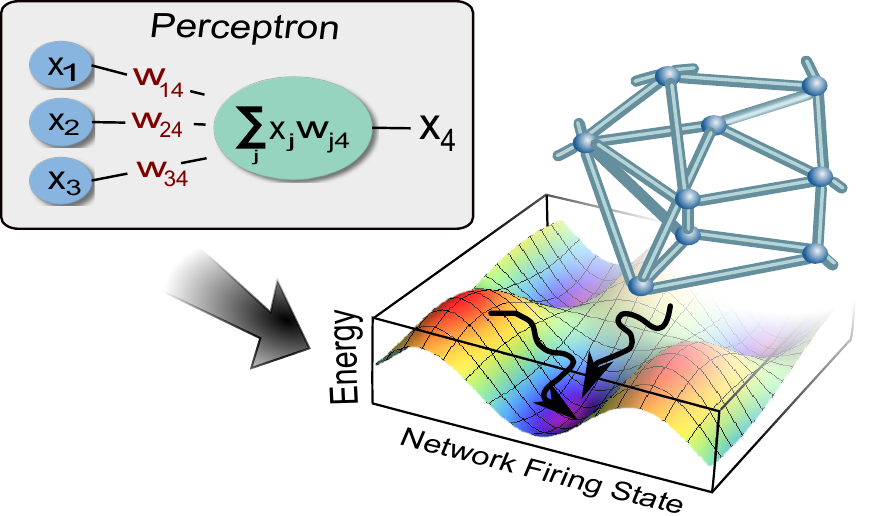} 
\caption{(Colour online) Illustration of a perceptron, a mathematical model of the neural activation mechanism with neurons $x_1,x_2,x_3, x_4$ and connection strengths $w_{14}, w_{24}, w_{34}$. In a recurrent Hopfield network, neurons with a perceptron activiation function are mutually connected as depicted in the graph structure. Such a network that stores firing patterns as minima of an energy function.} \label{neural}
\end{figure}

It turns out that important properties of the brain, such as the computation of incomplete or imperfect input, can be retrieved by the easiest model of a neuron possible, introduced by McCulloch and Pitts as early as 1943 \citep{mcculloch43}: an active neuron firing a sequence of action potentials in a given time is represented by a `1' while a resting neuron is represented by a `0'.  (for other types of neural networks, see \citep{rabinovich06}). The $N$ neurons of a neural network can thus be described by variables $x_i \in \{0,1\}, \; i = 1,...,N$. Each neuron $x_i$ is assigned to a characteristic threshold $\theta_i$. The biologically derived activation or updating function of a neuron $x_i$  is then given by 
\begin{equation} x_i =  \left\{   \begin{array}{l l}
		    1, & \quad \text{if} \; \sum\limits_{i \neq j=1}^N w_{ji} x_j \leq \theta_i,\\
   		    0, & \quad \text{else,} 
  		\end{array} \right . \label{update}\end{equation}
where the $w_{ij},\;  i,j = 1,...,N $ are real numbers denoting the strength of the connection between neuron $x_i$ and $x_j$. The according setup is called a `perceptron' (see Fig. \ref{neural}). The vector $(x_1,...,x_N)$ is called the `state' or firing pattern of the network. Initialising the network means to set each neuron to either $1$ or $0$. An update step of the global network state can either happen through a synchronous update of all neurons, through a chronological or random sequence of individual updates according to Eq. (\ref{update}).\\

One of the milestones in artificial neural network research was John J. Hopfield's 1982 publication on a network today widely known as `Hopfield network' \citep{hopfield82} in which the connection strengths fulfill
\[w_{ij} = w_{ji}, \qquad w_{ii} = 0. \]
Although of a simple setup, the Hopfield model  shows the powerful feature of associative memory. Associative memory is the ability to --  out of a number of stored firing network states -- retrieve the network state that is in the center of the dynamic basin of attraction for the input pattern. Hopfield networks thus store firing patterns as dynamic attractors. These dynamic attractors are minima of the Ising-type energy function
\begin{equation} E(x_1,...,x_N) = - \frac{1}{2} \sum\limits_{i=1}^{N} \sum\limits_{j=1}^{N} w_{ij} x_i x_j + \sum\limits_{i=1}^{N} \theta_{i} x_i. \label{energy} \end{equation}
A Hopfield network always inherits attractors from the nonlinearity of the updating process. The specific dynamics of a neural network are then solely defined by the choice of weights $w_{ij}$. The property $w_{ii} = 0$ makes sure that all attractors are stable states (as opposed to limit cycles of alternating states) \citep{rojas96}. After a finite number of updating steps, any initial state of the network will consequently end up in the `closest' attracting network state which is then reproduced by the updating process. An important implication is that neural networks based on a step activation function do manipulate information irreversibly due to its injectivity, i.e. a state of a neural network at timestep $t_{n-1}$ cannot be reconstructed from its state at timestep $t_n$. This might be different for other types of activation functions such as the sigmoid function.  It is also interesting to note that in conventional neural networks, the number of neural excitations (i.e. of neurons with the state `active') is not conserved, which is crucial for the attempt to construct quantum walks of excitations in Quantum Neural Network. \\

Approaches to QNNs  \citep{kak95, perus00, menneer95, zak98, gupta01, oliveira08, behrman99, li02} are mostly based on Hopfield-type neural networks. The basic idea of introducing quantum properties is to replace the McCulloch-Pitts neuron $x = \{0,1\}$ by a `qubit neuron' $\ket{x} $  of the two-dimensional Hilbert space $\mathcal{H}^2$ with basis $ \{\ket{0}, \ket{1}\}$. We propose to simply call this object a `quron'. The central property of a quron is that it can be in a superposition of its two firing states with the complex amplitudes $\alpha, \beta$
\begin{equation}\ket{x} = \alpha \ket{0} + \beta \ket{1}, \;\; |\alpha|^2 + |\beta|^2 = 1. \label{qubit} \end{equation} 
The state of a network with $N$ qurons thus becomes a quantum product state of the $2^N$-dimensional Hilbert space $\mathcal{H}^{2^N} =\mathcal{H}^2_{(1)} \otimes \dots \otimes  \mathcal{H}^2_{(N)}$
\[\ket{\psi} = \ket{x_1} \otimes \ket{x_2} \otimes \dots \otimes \ket{x_N} = \ket{x_1 x_2 \dots x_N}.\]
These are the firing states of a QNN on which a quantum walk will be constructed in the following section.

\section{Quantum walks between quantum neural network states}

The genius of the Hopfield model lies in the fact that operations on the neuron as a local unit  impose dynamics on the global network state. These global dynamics can be understood as a classical random walk between network states, 
\[(x_1,...,x_n)_{t_0} \rightarrow (x_1,...,x_n)_{t_1}  \rightarrow (x_1,...,x_n)_{t_2}  \rightarrow \hdots, \] beginning with the initial pattern and in each step jumping to the updated network state. After a finite number of steps, the chain reproduces the stable state serving as the output of the algorithm. \\

\begin{figure}[t]
  \centering
   \includegraphics[width = 0.3\textwidth]{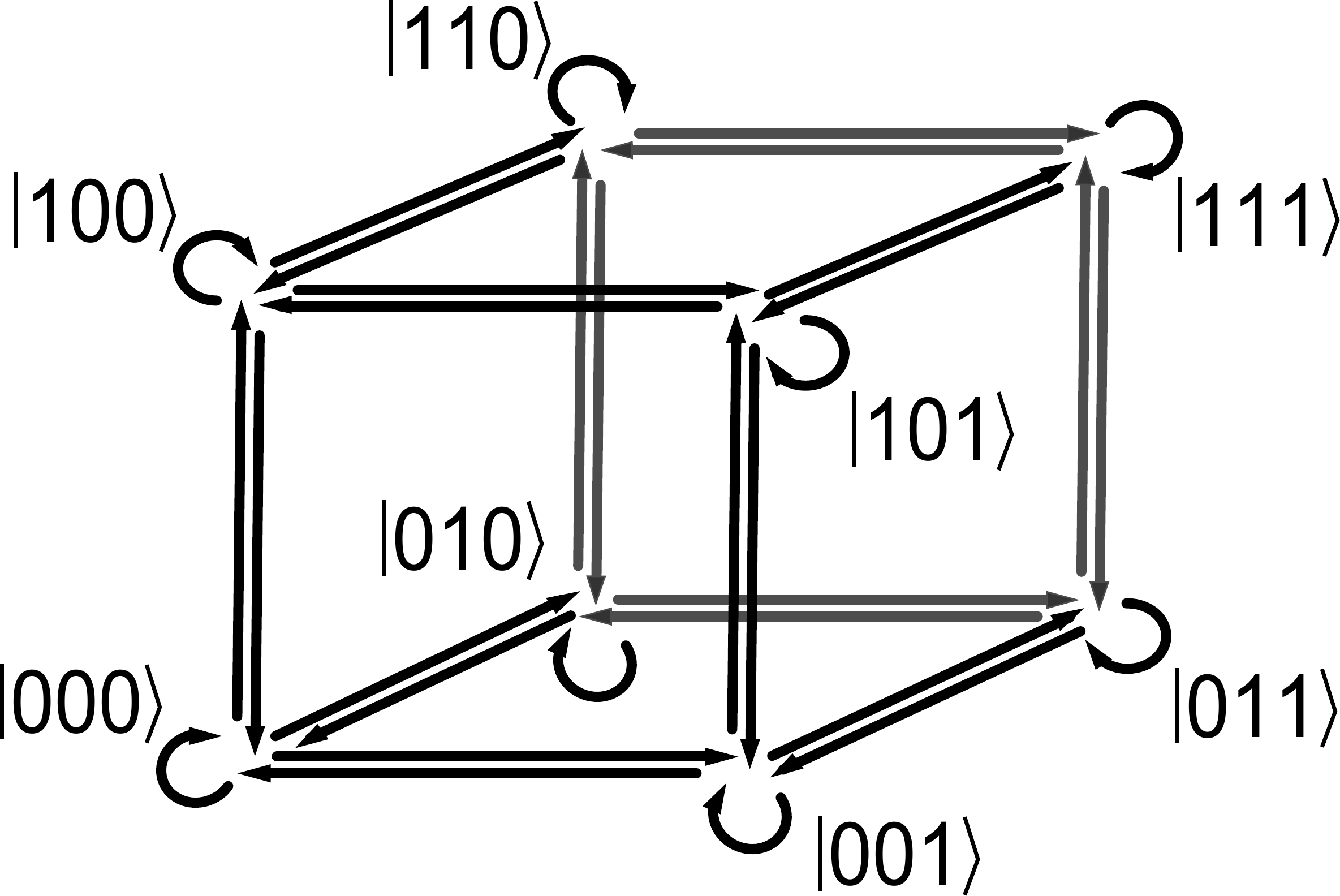}
\caption{The graph of a quantum walk on the firing states of a QNN has the structure of a hypercube, where the firing patterns sit at the corners.}
\label{HC}
\end{figure}

Likewise, a quantum walk on the firing states of a QNN can be defined as an evolution
\[\ket{\psi}_{t_0} \rightarrow \ket{\psi}_{t_1} \rightarrow \ket{\psi}_{t_2}\rightarrow \hdots,\]
following the laws of quantum mechanics. An important question is the structure of the graph underlying such a quantum walk. The vertices of the graph are given by binary strings denoting all possible firing patterns. The connectivity thus depends on the updating protocol. If all neurons are updated synchronously, each transition between firing patterns is theoretically possible and the graph is fully connected. We will concentrate on the more common case of updating single neurons at a time giving rise to the hypercube given in Fig. \ref{HC} for the $N = 3$ dimensional case. The hypercube connects binary strings that differ only in one bitflip (in other words, they have a Hamming distance of one \citep{hamming50}). We add self-connections of every site to account for updates that leave the firing pattern unchanged.\\

The remaining part of this article will mention a very intuitive way of implementing a quantum walk on a QNN by tossing a quantum coin to decide upon the updated state of each quron, and show why it fails to lead to a discrete, unitary quantum walk. It will then present a version of a Stochastic Quantum Walk that simulates an associative memory and discuss the results. 

\subsection{Why the most intuitive version of a discrete quantum walk fails}

A straight forward way to construct a QNN seems to be to replace the updating process of a neuron given in Eq. (\ref{update}) by a biased Hadamard-like transformation on a two-dimensional coin state $\ket{c}= \{\ket{0}_c, \ket{1}_c\}$ and to flip the state of the quron depending on the outcome of the coin as done in the discrete quantum walk. To retrieve nontrivial dynamics, we would want a biased coin leading to a superposition shown in Eq. (\ref{qubit}) in which the amplitudes $|\alpha|^2$ and $|\beta|^2$ encode a probability of the corresponding neuron to fire or to rest. This can be done by defining the firing probability $0\leq p_i \leq1$ of a neuron $x_i$ as
\begin{equation} 
p_i = \frac{\sum_j w_{ij} x_j + (N-1)}{2(N-1)}, 
\label{firingprob}
\end{equation}
and choosing
\[\alpha_i = \sqrt{1- p_i}, \qquad \beta_i = \pm \sqrt{ p_i} \]
The $\pm$ in front of $\beta$ is introduced to simulate the quantum properties of the Hadamard transformation or in other words, to introduce interference. The firing probability is nothing else than the normalised summed up signal coming from all input neurons to an output neuron. Since we sum over $N-1$ weighed neurons $w_{ij}x_j \in [-1,1]$, the incoming signal lies in the interval $[-(N-1), (N-1)]$. To obtain a positive value normalised to $[0,1]$ representing the probability, we consequently need to `shift' the signal to positive values and divide by the range of the interval, $2(N-1)$. Thus, if the incoming signal is strong, the probability for the neuron to become active is high, regardless of its prior state. Note that the thresholds $\theta_i$ of classical neurons are set to zero when dealing with qurons in the following. The updating process for quron $\ket{x_i}$ can consequently be formulated as a transformation 
\begin{multline} \hat{H} \ket{0}_c = \sqrt{1-p_i} \ket{0}_c +\sqrt{p_i}  \ket{1}_c, \\ \hat{H} \ket{1}_c =\sqrt{1-p_i} \ket{0}_c - \sqrt{p_i} \ket{1}_c. \label{neuroncoin} \end{multline}

This is slightly different from the biased Hadamard transformation used in biased quantum walks \citep{mackay02, kendon07, stefanak09, ribeiro04},
\begin{multline} \hat{H} \ket{0}_c = \sqrt{p_i} \ket{0}_c +\sqrt{1-p_i}  \ket{1}_c, \\ \hat{H} \ket{1}_c =\sqrt{1-p_i} \ket{0}_c - \sqrt{p_i} \ket{1}_c, \label{biasedcoin} \end{multline}
where the variable $p_i$ denotes the probability that the coin state flips its value, so that the biased Hadamard coin depends on the history of the coin state. This small difference leads to a problem in the implementation of the quantum walk proposed here, since the coin (Eq. \ref{neuroncoin})  is nonunitary. In fact, this property is not surprising since it stems from the dissipative dynamics of the Hopfield network, in which information of the former state of the neuron does not feed into the updating process (due to $w_{ii} = 0$). A direct application of coherent quantum walks onto neural dynamics is thus not trivial. As a conclusion, quantum walks including decoherence must be considered to incorporate dissipation.

\subsection{A Stochastic Quantum Walk on the hypercube}

We want to propose another type of quantum walk that is not derived from the neural updating mechanism, but still obtains the Hopfield network's dynamics of associative memory. Hence, our goal is to introduce a Stochastic Quantum Walk on a hypercube graph that ends up in one of two `attracting firing states' closer to the initial state in terms of Hamming distance.\\

\begin{figure}
         \centering
            \includegraphics[width=0.5\textwidth]{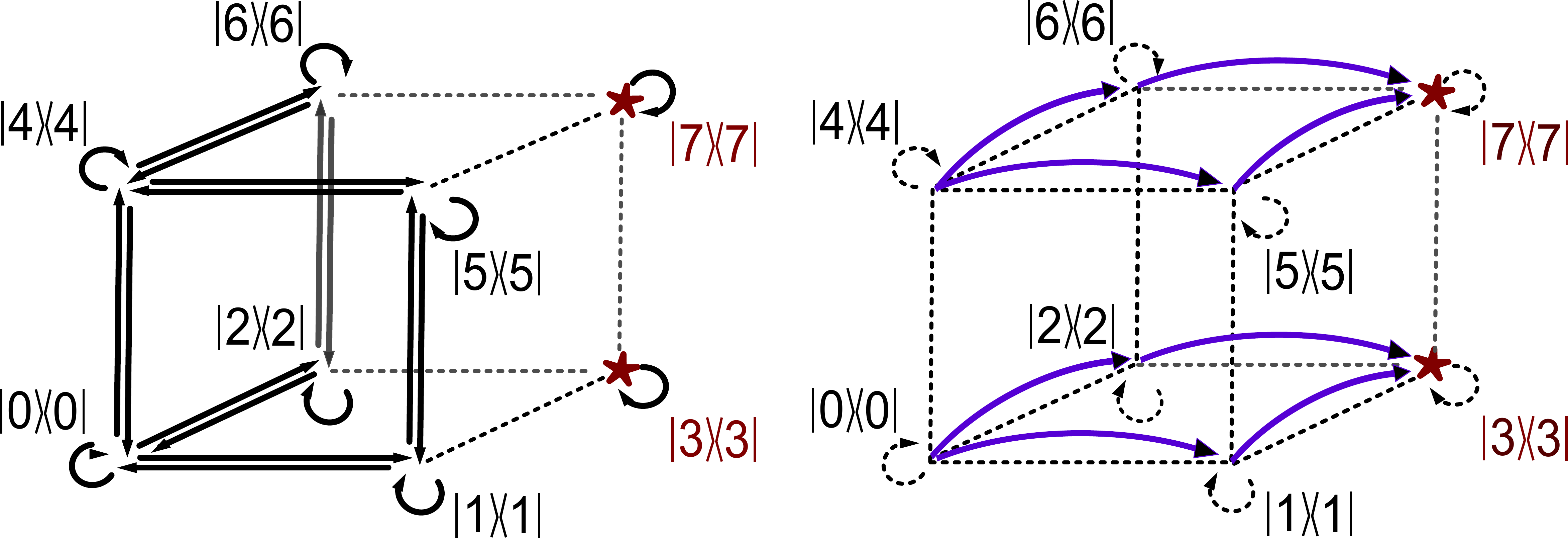}
              \caption{(Colour online) Illustration of the construction of the graph for the Stochastic Quantum Walk described in the text. The left figure represents the coherent part, using a Hamiltonian derived from the graph's adjacency matrix. The sinks ($\ket{7}\bra{7}$ and $\ket{3}\bra{3}$, in red) are isolated (dashed lines indicate no connection). The right figure represents the decoherent part of the walk. Lindblad jump operators (long arrows, in purple) simulate attraction by introducing a flow towards the sinks.}
  \label{sqwHC}
\end{figure}

The hypercube of dimension $N$ is given by a set of vertices $\mathcal{V}^{2^N}$ as `corners', representing the density matrices $\ket{x_1,...,x_N}_i\bra{x_1,...,x_N} = \ket{i}\bra{i}, i = 1,...,2^N$ (compare Fig. \ref{sqwHC}). The quantum state $\ket{x_1,...,x_N}_i$ is thereby the $i$-th basis state of a QNN of 2-level qurons as introduced above, and the shorthand $\ket{i}$ is used to reduce notation. In the hypercube, two vertices $\ket{i}\bra{i}$ and  $\ket{j}\bra{j}$ are connected by an edge if their respective network states differ by one quron's state or not at all, or in other words, the Hamming distance $d_H(i,j)$ between the two states $\ket{i}$ and $\ket{j}$ is one or zero. The hypercube graph's adjacency matrix is consequently given by
\[ H_{ij} = \left\{ 
   \begin{array}{l l}
    a_{ij}, & \quad \text{if } d_H(i,j) = 0,1 ,\\
    0, & \quad \text{else.} 
   \end{array}  \right. . \] 
For now we set $a_{ij}=1$ \footnote{The weights $a_{ij}$ can be chosen to be biased so that edges further away from both sinks get a lower value than those close to a sink. This gives another speed-up, especially in high dimensions.}. We introduce sinks by removing the edges leading to/from the vertices that represent the patterns we want to memorise. For simplicity we shall consider the example of only two `sink states', but this case can easily be generalised. Removing the edges is necessary since once the walker arrived at a sink, it is supposed to be trapped with no possibility to leave. Since in continuous quantum walks the adjacency matrix is equivalent to the hamiltonian, the graph structure needs to be undirected to ensure the hermiticity of the hamiltonian. This is why by removing the edges leading out of the sinks, we have to remove the edges leading to the sinks at the same time. The graph structure of the coherent part of the stochastic quantum walk is sketched in Fig. \ref{sqwHC} on the left.\\

The dissipative part of the GKLS master equation (\ref{lindblad}) can be written with the help of jump operators $L_k$. We use these jump operators to account for the `directed' part of the walk. Each edge $i\leftrightarrow j$ between vertices $ \ket{i}\bra{i}$ and $ \ket{j}\bra{j}$ is assigned with a jump operator $L_k =L_{ i\rightarrow j} =\ket{j}\bra{i} $ where $\ket{j}\bra{j}$ is the vertex closer or equal to a sink state. If both vertices sharing an edge have the same distance to a sink, no jump operator is attributed to that edge. This setup creates a flow to the sink states in the dissipative part and builds a `bridge' between the graph structure of the coherent part to the disconnected sinks. The graph structure of the dissipative part of the stochastic quantum walk is sketched in Fig. \ref{sqwHC} on the right. \\

The resulting master equation for the stochastic quantum walk with the two sink states $\ket{l}\bra{l},\ket{m}\bra{m}$  is then given by Eq. (\ref{lindblad}) with
\begin{align*} 
 H &= \sum \limits_{<i,j> \neq l,m}a_{ij} \ket{i}\bra{j},\\
 L_k  &= L_{ i\rightarrow j} =  \ket{j}\bra{i}, \\
\end{align*}
where $<i,j> = \{\ket{i}\bra{i},\ket{j}\bra{j} \in \mathcal{V}^{2^N} | d_H(i,j) =1 \}$ is a pair of connected vertices,  $i\rightarrow j =  \{\ket{i}\bra{i},\ket{j}\bra{j} \in \mathcal{V}^{2^N} |\;\; \mathrm{min}[d_H(j,l),d_H(j,m)]  \leq \mathrm{min}[d_H(i,l),d_H(i,m)]  \}$ denotes a pair of connected vertices in which  $\ket{j}\bra{j}$  is the vertex closer or equal to a sink state, and $a_{ij} = a_{ji}$.\\
\begin{figure}
         \centering
            \includegraphics[width=0.45\textwidth]{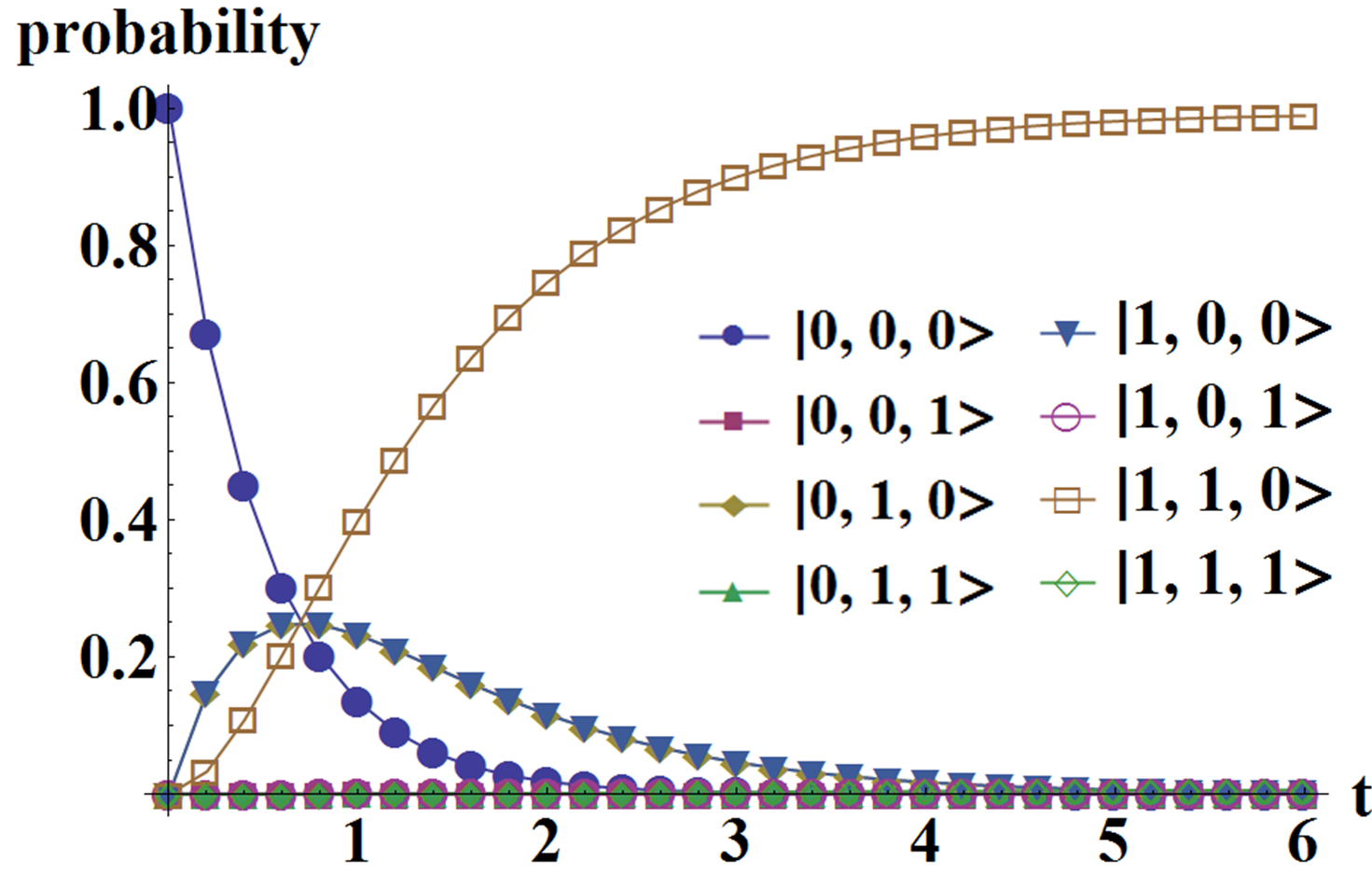}
              \caption{(Colour online) Example of the evolution of the QNN firing states' probabilities  in the Stochastic Quantum Walk on the hypercube of dimension $N =3$  as introduced here; with sinks at $\ket{101}\bra{101}$ and $\ket{111}\bra{111}$, initial state $\ket{000}\bra{000}$ and $\kappa = \gamma =1$. After only two time units, the walker has a high probability to be in the desired output state. Note that time is in inverse units of $\gamma$.}
  \label{HCplot1}
\end{figure}

\begin{figure}
         \centering
            \includegraphics[width=0.45\textwidth]{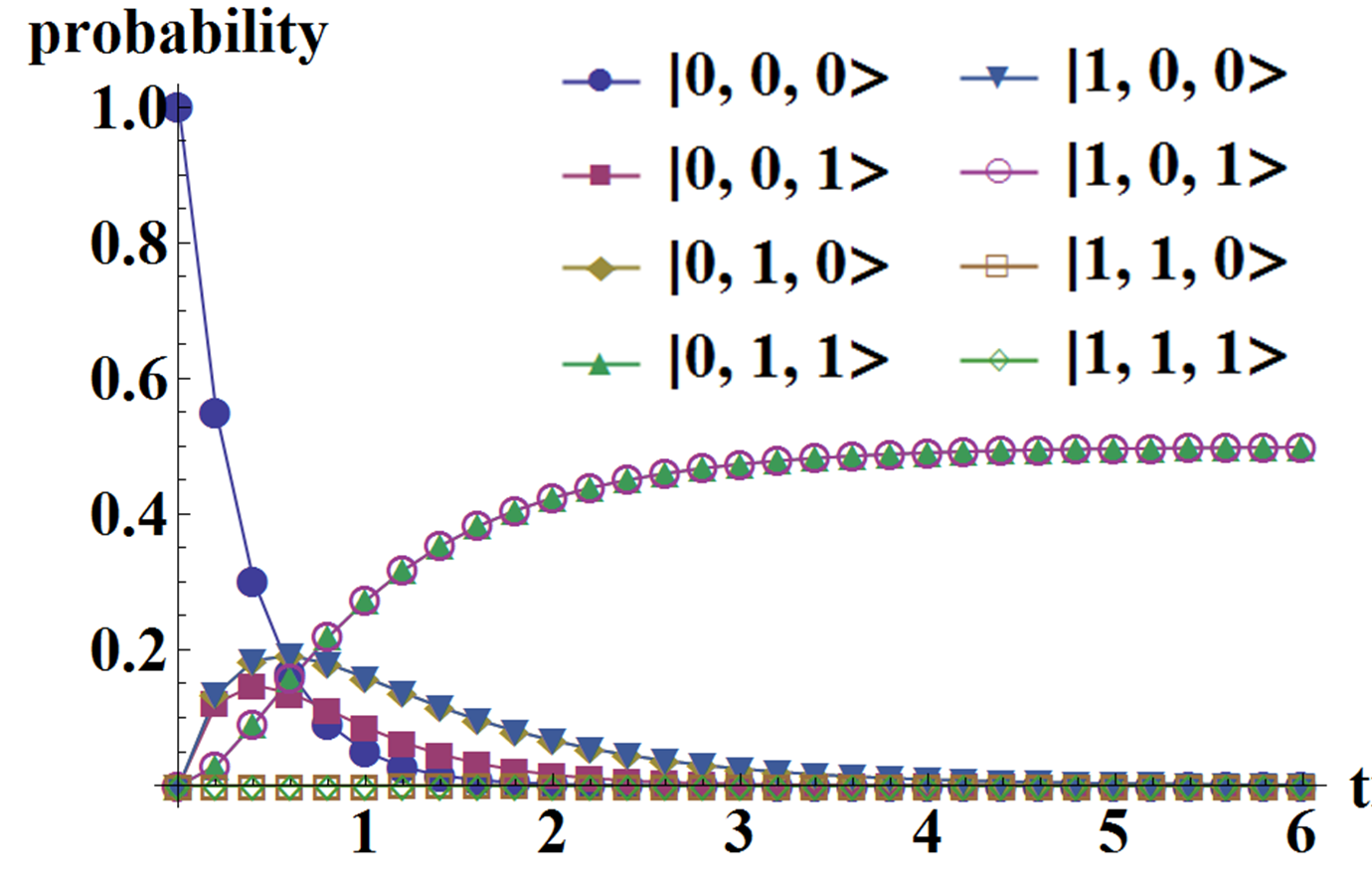}
              \caption{(Colour online) In this example, the sink states are given by $\ket{011}\bra{011}$ and $\ket{101}\bra{101}$, the initial state is $\ket{000}\bra{000}$ and $\kappa = \gamma =1$. Since both sink states have the same Hamming distance to the initial state, the walker has a probability of almost 0.5 to end up in either of the respective sink states. Note that time is in inverse units of $\gamma$. }
  \label{HCplot2}
\end{figure}

\subsection{Results}
The Stochastic Quantum Walk starts at the vertex representing the initial firing pattern and propagates over the hypercube. After a time evolution of the magnitude of several time units, the walk always finds the sink closest to the initial state in terms of Hamming distance with a probability of nearly $1$ (Fig. \ref{HCplot1}). The model consequently shows the basic neural network feature of associative memory. If the two sinks have the same distance, the output is an equal probability of both sink states (Fig. \ref{HCplot2}). This is an optimisation to a classical, deterministic associative memory which favours one of two states of an equal Hamming distance to the initial states. \\

\begin{figure}[t]
         \centering
                 \includegraphics[width=0.45\textwidth]{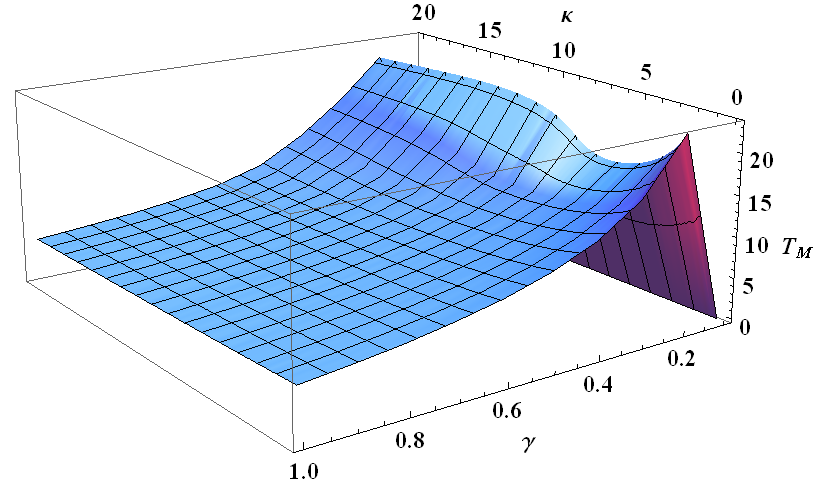}
              \caption{(Colour online) Mixing time $T_M$ in inverse units of $\gamma$ to reach a steady state in dependence of the parameters $\kappa$ and $\gamma$ in the Stochastic Quantum Walk on the $4$-dimensional hypercube with sink states $\ket{1011}\bra{1011}, \ket{1111}\bra{1111}$ and initial state $\ket{0000}\bra{0000}$. The speed of the algorithm is almost independent of the quantum contribution to the walk (represented by parameter $\kappa$). However, for low values of $\gamma$, there is an optimal value of $\kappa$ in terms of mixing time. Please note that if the Stochastic Quantum Walk did not show convergent behaviour, the mixing time was set to zero in order to indicate that the Quantum Associative memory did not retrieve the correct result.}
  \label{parHC}
\end{figure}

It turns out that the dynamics of the Stochastic Quantum Walk on the hypercube are mainly influenced by the incoherent part of Eq. (\ref{lindblad}). Figure \ref{parHC} shows the time until the average probability distribution reaches the correct stable state (mixing time) in dependence of the two parameters $\kappa$ and $ \gamma$ as given in Eq. (\ref{lindblad}). One can see that the time of reaching a stable distribution only depends on $\kappa$ for small values of $\gamma$, which denotes the coherent or quantum part in the stochastic walk. However, for values $\gamma < 1$, the quantum part can increase the speed of the walk by several time units. Since the quantum walk has shown to traverse the hypercube exponentially faster than a classical walk \citep{kempe02}, the contribution of the quantum speed-up might be larger in higher dimensions. Due to the exponential growth in the dimension of the Hamiltonian, simulations in dimensions $\geq 7$ require large computational resources. \\

\section{Conclusions}

This article studied some aspects of the application of quantum walks to Quantum Neural Networks. It was argued that a direct translation of the neural updating mechanism into a Hadamard-like transformation faces the problem of a nonunitary coin operator. This stems from the dissipative nature of the neural activation function and is symptomatic for the attempt to combine the attractor-like dynamics of neural networks and the linear, unitary dynamics of quantum objects. We concluded that decoherence needs to be introduced into the model. A Stochastic Quantum Walk on the hypercube was therefore constructed, and we could show its property of associative memory, an important feature of neural networks. Due to the low dependence on the quantum evolution or coherent part of the walk, this model is only under a limited perspective a candidate for a quantum walk on  the firing states of a Quantum Neural Network. However, these results can be seen as a first attempt in this direction and serve as an example of the application of quantum walks to obtain certain algorithmic dynamics of quantum systems. \\

There are other versions of quantum walks that might be worth investigating in order to overcome the flaws presented by Stochastic Quantum Walks and coined quantum walks. For example, there are different ways to introduce decoherence into the dynamics, such as projective measurements on the coin \citep{brundec03, kendon03, kendon07}. In fact, measurements have been proposed by several authors searching for a QNN model to account for the nonlinear updating process of neurons in a quantum regime \citep{kak95, perus00, menneer95, zak98}. Another interesting version of quantum walks are the recently developed Open Quantum Walks \citep{attal12, sinayskiy12, bauer13}. Based on the theory of open quantum systems, Open Quantum Walks describe a walker whose internal degrees of freedom are interacting with an environment and influencing the walker's external degree of freedom. The formalism shows a striking analogy to the updating function of neurons, giving the advantage that it does not require the global coherence of QNN states as in the walks on network states investigated here. Open Quantum Walks might consequently be a natural candidate when studying possible dynamics of Quantum Neural Networks.\\

The underlying idea to this paper was to use the formalism of quantum walks in order to find a dynamic evolution of a Quantum Neural Network that optimises the computational properties of classical neural networks \citep{schuld14}.  In a second step, the dynamic evolution could then be attempted to be attributed to physical processes, in the far picture possibly leading to a quantum model of biological neural networks. It can therefore be interesting to ask how the model can incorporate learning, a mechanism characteristic for neural networks. It is important to emphasize again that the nodes of the graph constructed in Fig. \ref{sqwHC} do not represent qurons, but entire firing states of a Quantum Neural Network and the edges $a_{ij}$ consequently do not correspond to the neural weights $w_{ij}, \; i,j= 1,...,N$. However,  similar to Hopfield networks `learning' a pattern by choosing appropriate weights that imprint the memory states into the energy function Eq. (\ref{energy}), choosing the connections $a_{ij}$  defines the dynamics of the Quantum Associative Memory model presented here. In both cases, learning is static, i.e. done by the initial construction of the network (or the graph). To get a quantum model that includes dynamic learning it would be a fruitful perspective to construct a quantum walk that simulates the above mentioned \textit{feed-forward neural networks}. Feed-forward networks are dynamically trained by so called backpropagation algorithms, where random initial weigths are repeatedly manipulated to minimise an error function comparing target outputs of a training set to real outputs calculated by the neural network \citep{rabinovich06}. Such a quantum walk would be required to reproduce feed-forward network's characteristica such as pattern recognition and could serve as an interesting continuation of the results found here.   \\

\section*{Acknowledgements} \textit{This work is based upon research supported by the South African Research Chair Initiative of the Department of Science and Technology and National Research Foundation.}

\end{document}